\documentstyle[12pt]{article}

\def\beqa{\begin{eqnarray}}
\def\eeqa{\end{eqnarray}}
\def\beq{\begin{equation}}
\def\eeq{\end{equation}}

\def\umu{^{\mu}}
\def\unu{^{\nu}}

\def\dmunu{_{\mu\nu}}
\def\ua{^{\alpha}}

\def\pa{\partial}

\def\bib#1{$^{\ref{#1}}$}

\let\alp=\alpha

\def\prl{{\it Phys. Rev. Lett.}\ }
\def\pl{{\it Phys. Lett.}\ }

\def\apj{{\it Ap. J.}\ }
\def\aj{{\it Astron. J.}\ }

\def\ie{{\it i.e. }}
\def\eg{{\it e.g. }}

\def\p{\phi}

\def\l{\cal L}
\begin{document}
\def\bib#1{[{\ref{#1}}]}
\begin{titlepage}
	 \title{Defocusing gravitational microlensing}

\author{{S. Capozziello$^{1,2}$, R. de Ritis$^{1,2}$, V.I. Mank'o$^{3,4}$,
A.A. Marino$^{1,2}$, G. Marmo$^{1,2}$}
\\ {\em $^{1}$Dipartimento di Scienze Fisiche, Universit\`{a} di Napoli,}
\\ {\em $^{2}$Istituto Nazionale di Fisica Nucleare, Sezione di Napoli,}\\
   {\em Mostra d'Oltremare pad. 19 I-80125 Napoli, Italy,}\\
 {\em $^{3}$Osservatorio Astronomico di Capodimonte,}\\
{\em Via Moiariello 16 I-80131 Napoli, Italy}\\
{\em $^{4}$Lebedev Physical Institute, Leninsky Pr., 53, Moscow 117924, 
Russia.}}

	      \date{}
	      \maketitle

  \begin{abstract}
We introduce the notion of defocusing gravitational lens considering
a MACHO located behind a light source with respect to an observer.
The consequence of defocusing effect is a temporal variability 
of star luminosity which produces a gap instead of a peak as tell--tale
signature in the light curve. 
General theory of (de)focusing rays (geodesics) in a gravitational field  
is presented. Furthermore, we give estimations of the mass of the lens
and the optical depth connected to such a phenomenon.

 \end{abstract}	      

\vspace{20. mm}
PACS: 95.30 Sf\\
e--mail address:\\
capozziello@axpna1.na.infn.it\\
deritis@axpna1.na.infn.it\\
manko@astrna.na.astro.it\\
marino@axpna1.na.infn.it\\
gimarmo@axpna1.na.infn.it

	      \vfill
	      \end{titlepage}

\section{\normalsize \bf Introduction}
Recently, gravitational lensing has become one of the most powerful tool in
astrophysics and cosmology to investigate 
the mass distributions and  the presence of dark matter in the universe
\bib{schneider},\bib{ehlers},\bib{peebles}.
In principle, it allows to estimate the gravitational mass of all large scale
structures, starting from galaxies to super cluster, and, in the specific
application called {\it microlensing}, it 
 can be used
to search for the so--called MACHOs {\it (Massive Astrophysical Compact
Halo Objects)} \bib{paczynski}, cosidered the most probable 
candidates for baryonic dark matter of Galaxy halo \bib{padmanabhan}
(however other possibilities are also 
explored \bib{gurevich1},\bib{gurevich2},\bib{gurevich3}).

Such objects may be considered as the main constituents of the dark halo of
spiral galaxies (in particular of our Galaxy) and, from theoretical constraints,
could have a very large mass range ($10^{-8}\div 10^{2}M_{\odot}$, so that they
could be little planets, big planets as Jupiter, brown dwarfs, or
massive black holes \bib{carr}).

The fundamental issue in this
approach is how lensing by a point--like mass can be detected.
Unless the lens is very massive ($M>10^{6}M_{\odot}$), 
the angular separation of  two
images (usually produced by a point lens) is too small to be resolved
(the angular separations of images
are of the order $\sim 10^{-6}$ arcsec, that is the reason for the term
{\it microlensing}).
However, when it is not possible to detect  multiple images, the
magnification can still be seen if the lens and the source move 
relatively to each other: this motion gives rise to 
a lensing--induced time variability
of the source luminosity\bib{chang}. 
 Such an effect
was first observed for the quasars QSO 2237+0305 and QSO 0957+561
\bib{irwin},\bib{schild}; so that we have to distinguish {\it galactic}
microlensing and {\it extragalactic} or {\it cosmological} microlensing.
In the first case, the light sources are stars and the angular separations
involved are $\sim 10^{-3}$arcsec, in the second case, the sources are
very distant quasars and the angular separations involved are
$\sim 10^{-6}$arcsec. In both cases the term "microlensing" is used.

The principle on which microlensing lies is quite simple. 
If the closest approach between a point mass 
lens and a source is equal or less  than $\theta_{E}$, the Einstein  angular
radius, the peak magnification in lensing--induced light curve 
corresponds to a brightness
enhancement (\eg $\sim 0.3$ magnitudes is a good number), 
which can be easily 
detected. The Einstein angular radius $\theta_{E}$, as we shall discuss below,
is a property of the system lens--source which furnishes the natural angular 
scale to describe the lensing geometry. In fact, for multiple imaging, 
it gives the typical angular separation among the single images; for
axisymmetric lens--source--observer systems, it gives the aperture of the
circular bright image, called {\it Einstein ring}
(the Einstein ring, as a geometric construction, can be defined in any
case, that is also if a luminous circular image is not produced). 
However, sources which are 
closer than $\theta_E$ to the optical axis experience strong lensing effect 
and are hardly magnified, sources which are located well outside of 
the Einstein ring are not very much magnified. In other words, 
for a lot of lens models, 
the Einstein ring represents the boundary between the zones where sources
are strongly magnified or multiply--imaged and those where they are softly
magnified or singly--magnified (actually the situation is very complicated
depending on caustics and Fermat's potential. For a detailed exposition
see for example \bib{ehlers}).

In order to detect microlensing,
the first proposal \bib{paczynski} was to monitor millions of stars 
in the
{\it Large Magellanic Cloud}
(LMC), or in the bulge of Galaxy in order to look for such magnifications. 
If enough events are detected, it should
be possible to map the distribution of (dark) stellar--mass objects in 
the halo  of 
Galaxy (due to the fact that LMC is near us and the halo of our galaxy 
is between) or between the Solar System and the bulge of Galaxy.
The two approaches involve some care in the selection of distances between
source and observer. In fact, the distance between the Sun and the center
of LMC is $\sim 55$Kpc while the distance between the Sun and the bulge
of Galaxy is $\sim 8.5$Kpc: this difference of size gives Einstein radii for 
the selected sources which could differ of about one order of magnitude.
Furthermore, the halo of Galaxy is supposed to extend of approximately
$\sim 50$Kpc so that the zone where MACHOs can pass is very large.
However, both approaches can be used for "galactic microlensing"
and, if we consider the Einstein radius $r_{E}\sim 1\div 10$AU, the 
distances of the source--lens--observer system $D\sim 1\div 50$Kpc,
and the velocities of passing MACHOs $v\sim 100\div 500$Km $s^{-1}$,
we are going to  give good numbers which can produce observable effects
\bib{paczynski},\bib{alcock}.

The biggest trouble of such a proposal was to distinguish
the intrinsic variable stars (which are very numerous in a normal galaxy)
and the lensing--induced variables. Fortunately, the light curves of lensed 
stars have certain features which allow to separate induced variability
from intrinsic variability (\eg the light curves are symmetric in time
and there are no chromatic effects since light deflection does not depends 
on wavelength; on the contrary, intrinsic variables have asymmetric light 
curves; furthermore, magnification produces chromatic effects
due to variability).

The expected time scale for microlensing--induced variations is given in 
terms of the typical angular scale $\theta_{E}$, the relative velocity $v$
between source and lens, and the distance of the observer to the lens 
$D_{ol}$:
\beq
\label{1}
\Delta t=\frac{D_{ol}\theta_{E}}{v}\,.
\eeq
If light curves are sampled with time intervals between  the hour and the
year, the mass range of MACHOs is $10^{-6}\div 10^{2} M_{\odot}$.
Such a time is directly connected to the so called {\it Shapiro delay}
\bib{shapiro} defined as
\beq
\label{2}
\Delta t=\int_{source}^{observer}\frac{2}{c^{3}}|{\bf \Phi}|dl\;,
\eeq
which gives the total time delay obtained by integrating over a light path 
modified by the Newtonian potential ${\bf \Phi}$ 
from the source to the observer.

We have to note  that we cannot get the mass of  MACHO $M$ directly
from (\ref{1}) since we have  the combination $M, D_{ol}, D_{ls}, D_{os}$ 
 (in the definition of $\theta_{E}$, see below) and $v$
from which we have to extrapolate $M$. This is a difficulty of the
theory since we need also accurate distance indicators  and accurate methods
to calculate  velocities of stars in the Galaxy.

Furthermore, we have to take into consideration the approximation 
we used: \ie the system 
lens--source is considered as formed by point--like objects.
In order to satisfy such an approximation, we need that
\beq
\label{small}
r_{lens}\ll r_{E}\;;
\eeq
so if we are considering galactic microlensing with 
$r_{E}\sim 1\div 10$AU, giants and supergiants stars are excluded since
they have sizes of this order of magnitude.
If we require, for example, $r_{lens}\leq 10^{-3}r_{E}$, this implies mass
densities of the order $\rho\geq 1$gr/cm$^{-3}$ and then low mass stars
(like those of Main Sequence),  brown and white dwarfs pass the requirement.
By using just Main Sequence stars as point--sources, as in most of the 
running experiments, we get a lower limit for the detectable lens mass 
coming from
\beq
\label{sun}
r_{E}\geq R_{\odot}\;;
\eeq
which implies for the mass
\beq
\label{sunmass}
M_{lens}\geq 10^{-6} M_{\odot}\,,
\eeq
being ${\displaystyle M=\frac{4}{3}\pi r^{3}\rho}$ the mass contained
in a sphere.
This means that if MACHOs are extremely light objects (\eg snowballs
with $M\ll 10^{-7}M_{\odot}$)
they cannot be detected.

Finally, we need also a statistical approach 
to microlensing since very little
information can be obtained by a single event; then we have to consider
some other questions essentially connected to:
$i)$ the expected rate of events; $ii)$ the distributions at different 
$\Delta t$
(the situation change if we look toward the bulge of the Galaxy
or toward the halo); $iii)$ the seasonal modulations
due to the Earth motions; $iv)$
the effects of binary (or multiple) stellar systems acting as 
 lenses or as sources; $v)$ the absorption effects which can drop
drastically the possibility of observations in certain zones of Galaxy.

Furthermore, the chance of seeing microlensing events depends on the 
{\it optical depth}, which is the probability that at any 
instant of time a given source is
within the angle $\theta_{E}$ of a lens. The optical depth is the integral 
over the number density $n(D_{ol})$ of lenses times the area enclosed by
the Einstein ring of each lens, \ie
\beq
\label{3}
\tau=\frac{1}{\Omega}\int dV n(D_{ol})\pi\theta^{2}_{E}\,,
\eeq
where  $dV$ is the volume of an infinitesimal spherical
shell with radius $D_{ol}$ which covers a solid angle $\Omega$. 
Eq.(\ref{3}) may assume a very simple form if the sources are distant
and compact objects, that is if sources and lenses have angular sizes
smaller than $\theta_{E}$.

Several groups are searching for MACHOs in the Galactic halo (by monitoring
stars in LMC) or in the Galactic bulge; among them we have:
MACHO \bib{alcock}, EROS \bib{aubourg}, OGLE \bib{udalski}, and DUO
\bib{alard}. So far, about 100 events have been detected and their number
is increasing rapidly. Most of them have been seen toward the Galactic bulge.
The majority of events have been caused by single lenses, but some of them
are due to binary lenses (which are distinguishable from single lens
events by characteristic double--peaked light curves).
However, we have, until now, few experimental data which can be considered
statistically relevant and allow to draw conclusions on the
physical properties of MACHOs like their mass.

What we want  to stress in our paper is the possibility of looking at the
microlensing from a different point of view. 

Microlensing is always  discussed for lenses which  focus light rays.
On the other hand, in optics, we know that there exists the opposite
effect if the refraction index of media is appropriately chosen
and if the relative positions of the source and the lens is changed 
with respect to the observer. That is, it seems natural to us 
 to ask the question whether there exist or not distributions of matter 
producing gravitational fields which deflect the light rays in a manner which
mimics defocusing lenses of standard optics. 

The wish to introduce and to study the notion of 
defocusing gravitational lens is motivated by the hypothesis that the 
microlensing events with luminosity peak may be accompained by the
existence of events with valley in the luminosity curve. This inverse 
phenomenon, in principle, may be understood if the relative positions of
MACHOs (lenses), stars (standard sources) and the observer
are taken into  consideration. Usually, the studied situation is that a
MACHO is between the source and the observer. The  emitted rays  by the
source are slightly curved in the direction of the observer and such a
fact produces the effect of luminosity magnification. Obviously, the
opposite situation is statistically as probable as previous one when a
MACHO is located behind the source with respect to the observer.  Then, the 
source rays are slightly curved out of the observer direction 
which may detect a
decreasing  luminosity. In other words, when a MACHO moves behind the 
source, its gravitational field produces a  
defocusing action. 

The aim of this paper is to deal with both defocusing and focusing effects
using  a unique  model. Thus we will describe the situation
in which a "beam" of initial geodesics is squeezed by the gravitational field
(or focused by the field) and produces an increasing of detected luminosity 
as well as the opposite case, when the initial ray beam radiated by source
is enlarged by the gravitational field, that is when 
we have  a decreasing (or defocusing) of detected  luminosity.
This general discussion may be precisely formulated using the equations for 
geodesics in a generic Schwarzschild gravitational field.
  
An important  implication of our approach is 
that in this way it is possible to improve the number of detected events
of MACHOs  which produces microlensing effects since 
it would be taken into account
valleys as well as peaks
in light curves.
We  study both phenomena (focusing and defocusing) 
 describing 
light trajectories starting from sources that, at the beginning, are
straight lines  and, passing near the 
gravitational field of a deflecting point mass (MACHO),
differ
from straight lines becoming, for example, a bundle of hyperbolic--like
geodesics converging or diverging toward the observer.

The paper is organized as follows.
In Sect.2, we deal with 
the generalities of point mass lenses obtaining the characteristic Einstein
radius, the magnification, and introducing the concept of optical depth. 
Sect.3, is devoted to the discussion of 
geodesics in a point--like gravitational field
giving the trajectories of (de)focused light rays.
In Sect.4, we discuss how (de)focusing detection can be 
realized considering the deviation angles of ray paths. 
We apply the obtained results in Sect.5 by calculating the mass of the lens
(MACHO) by (de)focusing and the optical depth, that is
the probability to get significant lensing for randomly located compact sources.
We  draw conclusions in Sect.6.

\section{\normalsize \bf Generalities on point mass lens model}
Gravitational lensing essentially consists in the deflection of light
 in gravitational fields as predicted by the
theory of General Relativity \bib{papapetrou}.
For small deflection angles and weak gravitational fields, which
are the regimes of practical interest, the true position of a light source
on the sky with respect to the position of its image(s) can be defined
by the lens equation
\beq
\label{lens}
\vec{\theta}-\vec{\theta}_s=
-\left(\frac{D_{ls}}{D_{os}}\right)\vec{{\tilde{\alp}}}(\vec{\theta})=
\vec{\alp}\,,
\eeq
where $\vec{\theta}$ defines the position(s) of image(s) with respect to the 
optical axis, $\vec{\theta}_s$ the position of the source, and
$\vec{{\tilde{\alp}}}$ is the displacement angle. $D_{ls}$ and $D_{os}$
are respectively the distances between the lens and the source and the 
distance between the observer and the source.

We have to note that a given image position always corresponds to a 
specific source position whereas a given source position may correspond
to several distinct image positions. Then we can have multiply imaged
sources. 
In the case of a point mass lens, as we will precisely show  
in the next section, the deflection angle is given by
\beq
\label{deflection}
\alp=\frac{4 GM}{c^{2}r_{0}}\,,
\eeq
where $M$ is the mass of  deflecting body; $r_{0}$ is 
  the minimal distance between the passing light
ray and the deflecting body \bib{ehlers},\bib{papapetrou}.

For point mass lenses, the geometry of the system is simplified and we do not 
need the full vector equation (\ref{lens}).
By writing  $r_{0}=\theta D_{ol}$, the lens equation
for a point--mass lens assumes the form
\beq
\label{point}
\alp=
\left(\frac{4GM}{c^{2}\theta}\right)\left(\frac{D_{ls}}{D_{os}D_{ol}}\right)=
\theta-\theta_{s}\,,
\eeq
which can be rewritten as:
\beq
\label{0.5}
\theta^{2}-\theta_{s}\theta-\theta_{E}^{2}=0\,,
\eeq
where
\beq
\label{0.3}
\theta_{E}=\sqrt{\frac{4GM(\leq r_{E})D_{ls}}{c^{2}D_{ol}D_{os}}}\,,
\eeq
is the Einstein angle which corresponds to the Einstein radius
\beq
\label{radius}
r_{E}=\theta_{E}D_{ol}\;,
\eeq
already introduced. 

We see that it strictly
depends on the distances involved and the mass of
the deflector. The symbol $M(\leq r_{E})$ means that the mass of the lens
has to be contained inside a sphere whose radius is the Einstein one.

Before solving and discussing the algebraic Eq.(\ref{0.5}), we have to spend 
some words on an important parameter connected to the lensing effect, 
 the {\it magnification}. In fact, gravitational lensing preserves
the surface brightness of a source and then the ratio (magnification)
between the solid angle $d\Omega_{i}$ covered by the lensed image and that of
the unlensed source $d\Omega_{s}$ gives the flux amplification
due to the lensing; this is given by the Jacobian of the transformation
matrix between the source and the image(s), that is
\beq
\label{0.8}
\mu=\frac{d\Omega_{i}}{d\Omega_{s}}=
\left|\mbox{det}\left(\frac{\pa\vec{\theta}_{s}}{\pa\vec{\theta}_{i}}\right)
\right|^{-1}\;.
\eeq
If there are more than one images of a source, the total magnification
is the sum of all image magnifications. 
Considering, as we are actually doing, a gravitational point mass lens 
system which is
axially symmetric with respect to the line--of--sight, we can use for 
deflection the scalar angle (\ref{deflection}) and apply Gauss's law
for the total flux. The light deflection reduces to a one--dimensional problem
and Eq.(\ref{0.8}) becomes
\beq
\label{0.9}
\mu=\frac{\theta_{i}d\theta_{i}}{\theta_{s}d\theta_{s}}\,,
\eeq
which is easily appliable \bib{ehlers}.

Let us now solve Eq.(\ref{0.5}). We get
\beq
\label{0.6}
\theta_{\pm}=\frac{\theta_{s}}{2}\pm\sqrt{\frac{\theta_{s}^{2}}{4}+
\theta_{E}^{2}}\;;
\eeq
from which we see that
\beq
\label{0.7}
\theta_{s}=0;\;\;\;\;\longrightarrow\;\;\;\;\theta_{\pm}=\pm\theta_{E}\,.
\eeq
Eqs.(\ref{0.6}), (\ref{0.7}) tell us that we have to expect at least
two images from the same source which lie on the same plane of
the source. In microlensing, as we discussed,
it is difficult to separate them and the effect results in a luminosity
enhancement of source.
The magnification corresponding
to Eq.(\ref{0.6}) is
\beq
\label{0.11}
\mu_{\pm}=\left[1-\left(\frac{\theta_{E}}{\theta_{\pm}}\right)^{4}
\right]^{-1}\,,
\eeq
which tells us that if $\theta_{s}$ is zero, the magnification becomes
singular; physically, this means that when the optical system
source--lens--observer is aligned, we can get a huge magnification.
The total amplification due to both images is
\beq
\label{0.12}
\mu=|\mu_{-}|+|\mu_{+}|=\frac{\chi^{2}+2}{\chi\sqrt{\chi^{2}+4}}\;;
\eeq
where
\beq
\chi=\frac{\theta_{s}}{\theta_{E}}\;.
\eeq
Immediately, we see that 
\beq
\label{0.13}
\theta_{s}\leq\theta_{E}\;\;\;\longrightarrow\;\;\;\mu\geq 1.34\,,
\eeq
which is the condition on the magnification inside the Einstein ring:
we have that a magnification $\mu\sim 1.34$ corresponds to
a magnitude enhancement of $\Delta m\sim 0.32$ as required in microlensing
experiments. In other words, we can say that when the true position of a light
source lies inside the Einstein ring, the total magnification of the two
images that it yields amounts to $\mu\geq 1.34$. This means that the angular 
cross
section for having significant lensing (\ie $\mu\sim 1.34$ and 
$\Delta m \sim 0.32$),  is equal to $\pi\theta_{E}^{2}$, which from
(\ref{0.3}), is proportional to the mass $M$ of the deflector and to
the ratio of the distances involved.
Such considerations allow to calculate one of the most useful quantities
for lensing detection: the optical depth.
Let us consider the case of randomly distributed point--mass lenses:
it is possible to estimate the frequency of significant gravitational lensing 
events from the observations of distant compact sources, that is we are
considering optical systems where the involved angular sizes  are 
much smaller than $\theta_{E}$. In this situation, the magnification of a 
compact source is equal or greater than $1.34$ 
(since $\theta_{s}<\theta_{E}$) and the probability $P$ to have significant
lensing for a randomly located compact source at a distance $D_{os}$
is given by
\beq
\label{0.14}
P=\frac{\pi\theta_{E}^{2}}{4\pi}=\left(\frac{D_{ls}}{D_{os}D_{ol}}\right)
\left(\frac{GM}{c^{2}}\right)\,,
\eeq
where we have used the definition (\ref{0.3}). 
Such a probability is linear in 
the mass $M$ of deflector so that it holds also when several 
point--mass lenses are
acting since the masses can be summed up. Assuming a constant density 
for the lens(es) and a static background (this last assumption
surely holds for galactic distances), averaging   on the distances
$D_{ls}, D_{ol}, D_{os}$, the probability (\ref{0.14}) can be interpreted as 
the  optical depth $\tau$  for lensing 
\bib{ellis},\bib{harwit},\bib{refsdal},\bib{press}. 
\beq
\label{optical}
P=\tau=-\left(\frac{D_{ls}}{D_{os}}\right)\frac{U}{c^{2}}\,,
\eeq
where
\beq
\label{newton}
U=-\frac{GM}{D_{ol}}\,,
\eeq
is the Newton potential due to the lens and measured by the observer.
If inside the Einstein ring there are several deflecting bodies,  
$U$ is their 
additive Newtonian potential.
In other words, $\tau$ corresponds to the fraction of sky covered by the 
Einstein ring. Due to the fact that the deflecting masses change the
path of light rays,  the observer will detect different luminosities
for a given source when the deflector is present and when it is not
present: then,  the optical depth will depend on such a
relative luminosity change.

In Sect. 5, we discuss some quantitative estimations  of such 
quantities in connection to 
lenses which focus or defocus light rays.

\section{\normalsize \bf Geodesics and light ray paths 
in a gravitational field}

Before considering  how to realize (de)focusing, it is useful to
discuss the motion
 of light ray paths in a  gravitational field since this fact allows to
derive the luminosity variations due to the presence of light deflecting
gravitational masses (in our case MACHOs).

In General Relativity, light rays 
move along geodesics 
\bib{papapetrou},\bib{wald}.
This fact means that, given the line element, 
$ds^{2}=g\dmunu dx\umu dx\unu$, 
they have to satisfy the  equations 
\beq
\label{2.2}
\frac{d^{2}x\ua}{ds^{2}}+\Gamma\ua\dmunu\frac{dx\umu}{ds}\frac{dx\unu}{ds}=0\,,
\eeq
where
\beq
\Gamma\ua\dmunu=\frac{1}{2}g^{\alpha\delta}(g_{\delta\mu,\nu}
+g_{\delta\nu,\mu}-g_{\mu\nu,\delta})\,,
\eeq
are the Christoffel symbols
and $s$ is the parameter chosen along the trajectory.

Here, we are in the 
geometric optic approximation so that the light propagates 
as rays and we have not
to take into consideration chromatic effects.
For weak gravitational fields (usually considered 
in gravitational  lensing effects), 
the metric tensor components can be expressed in terms of
 Newton gravitational potential $\Phi$ as 
${\displaystyle
g_{00}\simeq 1+2\frac{ \Phi ({\bf r})}{c^{2}}}$ and
${\displaystyle g_{ik}\simeq -\delta_{ik}\left(1-2\frac{\Phi({\bf 
r})}{c^{2}}\right)}\,,$
  the approximation $\Phi/c^{2}\ll 1$ holds. 

As it is well known
\bib{ehlers}, 
 a gravitational field has the same effect of a
medium (different from vacuum) in which light rays propagates and
the Fermat principle holds.
The 
refraction index $n$  can be expressed in terms of the 
gravitational potential $\Phi({\bf r})$ produced by some matter distribution
\bib{ehlers}), 
that is
\beq
n=1-\frac{2\Phi(r)}{c^{2}}\,.
\eeq

If the rays pass near a body of mass $M$, they will undergo the action
of a Schwarzschild gravitational field described by the metric element
\beq
\label{2.3}
ds^{2}=\left(1-\frac{R_s}{r}\right)c^{2}t^{2}-
\frac{dr^{2}}{{\left(1-\frac{R_s}{r}\right)}}-
r^{2}\left(d\theta^{2}+\sin^{2}\theta d\phi^{2}\right)\,,
\eeq
where
\beq
\label{schwarzschild}
R_{s}=\frac{2MG}{c^{2}}\,,
\eeq
is the Schwarzschild radius  which tells us where
the metric becomes singular.
We get the trajectories of light rays  by the
Lagrangian
\beq
\label{lagrangian}
{\l}=\left(1-\frac{R_s}{r}\right)(\dot{x}{^{0}}){^{2}}-
\frac{\dot{r}^{2}}{{\left(1-\frac{R_s}{r}\right)}}-
r^{2}\left(\dot{\theta}^{2}+\sin^{2}\theta \dot{\phi}^{2}\right)\,,
\eeq
obtained by line element (\ref{2.3}). 
The derivative is with respect to $s$. Its
Euler--Lagrange equations 
 are nothing else 
but the geodesics equations (\ref{2.2}).

They gives the condition for the planar motion,
the conservation of angular momentum
${\displaystyle r^{2}\dot{\phi}=k}$, and
 the conservation of energy
${\displaystyle
\left(1-\frac{R_s}{r}\right)\dot{{x^{0}}}=E}$. $E$ and $k$ are integration 
constants.

Substituting such results into the equation for $r$,
we get the equation for the 3--dimensional trajectories of rays
as function of $\phi$, that is
\beq
\label{2.8}
u''+u=\frac{3}{2}R_s u^{2}\,,\;\;\;\;\mbox{where}\;\;\;\;u=\frac{1}{r}\,.
\eeq
The prime indicates the derivative with respect to $\phi$.
The {\it rhs} of (\ref{2.8}) gives rise to the relativistic effects
of deflection of light.
In fact, a particular solution of (\ref{2.8}) is $u'=0$ 
which tells us that photons
can stay in a circular orbit having the radius $r={\displaystyle 
\frac{3}{2} R_{s}}$.

The general solution of (\ref{2.8}) is an elliptical integral \bib{byrd}
of the form
\beq
\label{elliptical}
{\cal K}(u,A_{0})=\int \frac{du}{\sqrt{R_s u^{3}-u^{2}+A_{0}}}=\p-\p_{0}\,,
\eeq
where $A_{0}$ and $\p_{0}$ are integration constants. The integral 
(\ref{elliptical}) cannot be inverted unless we approximate it or
choose particular initial conditions.

However, we are considering 
weak gravitational fields which act as 
perturbations on the straight light ray trajectories, 
\ie we are in the regime
\beq
\label{2.9}
\frac{3R_s u^{2}}{2u}=\frac{3}{2}\frac{R_{s}}{r}\ll 1\,.
\eeq
Condition (\ref{2.9}) means that  the light rays pass 
 far from the critical
radius $R_{s}$ where the gravitational field becomes singular
(this fact is quite obvious since for the usual astrophysical bodies we have
$R_{s}\ll R_{0}$ where $R_{0}$ is the surface radius).

If the gravitational source is absent (\ie $M=0$), the general solution of
(\ref{2.8}) is
\beq
\label{straight}
\tilde{u}=\frac{1}{r}=\frac{1}{r_{0}}\cos(\phi-\phi_{0})\,,\;\;\;\;\;\;\;
-\frac{\pi}{2}\leq\p-\p_0\leq\frac{\pi}{2}\,,
\eeq
which is a straight line in polar coordinates.
The parameters
$r_{0}$ and $\phi_{0}$ are the initial data of the problem; $r_{0}$
 is the distance of the line from
the origin of coordinates, $\p_{0}$ is a given angle which tells us how much
the line is tilted with respect to the polar axis.  

If condition (\ref{2.9}) holds, the {\it rhs} of Eq.(\ref{2.8}) can be 
treated as a small perturbation and then we can search for solutions of the 
form 
\beq
u=\tilde{u}+\epsilon u_{1}\,,
\eeq  
where
\beq
\epsilon=\frac{3R_s}{2r_{0}}\ll 1\,.
\eeq
Eq.(\ref{2.8}) becomes
\beq
u_{1}''+u_{1}=\frac{1}{r_{0}}\cos^{2}(\p-\p_{0})\,,
\eeq
which admits the solution
\beq
u_{1}=\frac{1}{3r_{0}}\left[2-\cos^{2}(\p-\p_{0})\right]+A_{1}\cos(\p-\p_{0})+
A_{2}\sin(\p-\p_{0})\,.
\eeq
The term in $A_{1}$  can be interpreted  as a redefinition
of $r_{0}$, the term in $A_2$ as the consideration of the
straight line perpendicular to (\ref{straight}): they can be both absorbed
by redefining the initial data. Finally, in the limit (\ref{2.9}), 
the solution of (\ref{2.8}) is 
\beq
\label{2.10}
u(\phi)=\frac{1}{r}=\frac{1}{r_{0}}\left\{\cos(\phi-\phi_{0})+
\frac{R_s}{2r_{0}}\left[2-\cos^{2}(\phi-\phi_{0})\right]\right\}\,,
\eeq
which is nothing else but a straight line
corrected by a
hyperbolic--like term in polar coordinates; the deflecting mass is 
set at the origin of reference frame. The amount of  deviation
from the rectilinear behaviour depends on the ratio 
${\displaystyle \frac{R_s}{r_{0}} }$, that is on the mass $M$ of the 
gravitational source and on the  parameter $r_{0}$.

Conversely, passing  to Cartesian coordinates
\beq
\label{2.16}
x=r\cos\phi\,,\;\;\;\;\;\;y=r\sin\phi\;;
\eeq
Eq.(\ref{2.10}) becomes
\beq
\label{2.17}
r_{0}=Ax+By+\left(\frac{R_s}{r_{0}}\right)\sqrt{x^{2}+y^{2}}-
\left(\frac{R_{s}}{2r_{0}}\right)\frac{(Ax+By)^{2}}{\sqrt{x^{2}+y^{2}}}\;,
\eeq
where 
\beq
\label{2.20}
A=\cos\phi_{0}\,,\;\;\;\;B=\sin\phi_{0}\;.
\eeq
The formula (\ref{2.17}) will be useful for the discussion below.

Let us consider now the limit
$r\rightarrow\infty$. 
Eq.(\ref{2.10}) is an algebraic equation for $\cos(\p-\p_{0})$ 
whose solutions are
\beq
\cos(\phi-\phi_{0})=
\left[\frac{r_{0}}{R_s}\pm\sqrt{\left(\frac{r_{0}}{R_s}\right)^{2}+2}\right]\,.
\eeq
Neglecting the positive sign solution which is without meaning, 
approximating
the term under the square root, 
 we get
\beq
\label{2.11}
\cos(\phi-\p_{0})\simeq -\frac{2MG}{c^{2}r_{0}}\,,
\eeq
which indicates how the presence of gravitational field $(M\neq 0)$ deviates
the rays from the straight line direction. If
$M=0$ or $r_{0}\rightarrow\infty$ (that is in absence of gravitational field
or when  $r_{0}$ is very large), we have
\beq
\label{2.12}
\cos(\phi-\p_{0})=0\,,\;\;\;\;\;\;\;\;\phi-\p_{0}=\pm\frac{\pi}{2}\;,
\eeq
while, if the gravitational field is weak in the limit 
$r\rightarrow\infty$, we have
\beq
\label{2.13}
\phi-\p_{0}=\pm\left(\frac{\pi}{2}+\delta\right)\;;
\eeq
from which, by substituting into (\ref{2.10}), we get
${\displaystyle \sin\delta\simeq \delta=\frac{R_s}{r_{0}}}$
being $\delta$ small. The total amount of ray deviation gives the standard
result
\beq
2\delta\simeq\frac{4MG}{c^{2}r_{0}}\,,
\eeq
which is  the deflection angle due to a point mass 
acting as a gravitational lens (see Eq.(\ref{deflection})).
If $r_{0}\simeq R_{\odot}$ and $M\sim M_{\odot}$, we get the classical
Eddington--Einstein result of $\delta\sim 1.75''$.

\section{\normalsize \bf (De)focusing and luminosity variation of the source}

Now, taking in mind the results of previous section,
 we want to obtain the general formula describing the 
 variation of   luminosity of a radiation source
in the sky induced by  a gravitational microlensing effect.
We will show that such a variation is due to the change of
 direction of  light rays (geodesics)  which move in a given  
nonstationary matter distribution and the effect is observable for
a time $\Delta t$ given by (\ref{1}). In other words, we are
supposing that a given
background metric $g^{(1)}\dmunu$ is modified by a passing heavy body
(a MACHO) which locally perturbs it so that we have to consider  a new metric
$g^{(2)}\dmunu$. The effect of such a background change is a deviation in the
direction of geodesics which can result, as above 
shown, a bundle of hyperbolic--like curves instead of 
a bundle of straight lines.

We will consider the following two cases:
 the observable variation
of source luminosity is due  to a microlensing focusing effect
by  a gravitational lens and by a microlensing
defocusing effect. 
In the first case, a MACHO  is between the source
and the observer producing focusing; in the second case, a MACHO 
 is behind the source and light rays are defocused
toward the observer. In the first case, the observer detects an increasing
 luminosity, in the second case, he detects a decreasing one.
The problem can be easily formulated by a geometric model
in which, given a reference frame, 
we assign the position of the light source and the position of 
the detector (a telescope)
in a  background metric $g^{(1)}\dmunu$. Then we calculate
the geodesics which give the light--ray paths. Then, considering a 
MACHO passing
between the source and the observer or behind the source (with respect to the
observer), the metric becomes $g^{(2)}\dmunu$ and the geodesics will change
giving focusing or defocusing of light rays. 

Let us  start by choosing a system of Cartesian
coordinates. We put the source in
\beq
\label{3.2}
(x_{S},\, y_{S})=(-a,\,0)\,.
\eeq
and the telescope   in
\beq
\label{3.3}
(x_{T},\, y_{T})=(R,\,h)\,,
\eeq
as shown in Fig.1. 
There exists a unique light path (a unique geodesic) 
which intersects the source and the
upper limit of telescope aperture (see again Fig.1). 

Let us now suppose that, due to a redistribution of matter,
the metric becomes
$g^{(2)}\dmunu$,
(the simplest case, as we said, is to consider  a MACHO passing 
 either between the source
and the observer or behind the source). 
This fact modifies the structure of geodesic bundle  from the source to 
the observer. Schematically, we have  a new geodesic between the source
and the upper limit of the aperture of telescope as shown
in Figs.2 and 3.

The ray which reachs the upper limit of  telescope 
in the  metric $g^{(1)}\dmunu$ is emitted at the angle $\alpha_{1}$ while 
it is  emitted at the angle $\alpha_{2}$ in the metric
$g^{(2)}\dmunu$.

In the first case,  geodesic is given by a function
$y_{1}(x)$ in Cartesian coordinates; in the second one by 
a function $y_{2}(x)$. The angles
$\alp_{1}$ and $\alp_{2}$ are given by the derivatives
\beq
\label{3.5}
\tan\alp_{1}=\frac{dy_{1}}{dx}\left|_{(-a;0)}\right.\,,\;\;\;\;\; 
\tan\alp_{2}=\frac{dy_{2}}{dx}\left|_{(-a;0)}\right.\,,
\eeq
calculated in the coordinates of the source.
Since the distances are very large the angles are small, so we have  
\beq
\label{3.6}
R\gg h\,,\;\;\;\mbox{and}\;\;\;\;R+|a|\gg h\,,
\eeq
and then
\beq
\label{angle}
\alp_{1}\simeq\frac{dy_{1}}{dx}\left|_{(-a;0)}\right.\,,\;\;\;\;\; 
\alp_{2}\simeq\frac{dy_{2}}{dx}\left|_{(-a;0)}\right.\,.
\eeq
In  Fig.2, we   show a focusing situation where metric is changing so that
 more light rays reach the telescope than in initial
metric (Fig.1). It means that the luminosity of the source 
detected by the telescope 
becomes larger. In Fig.3, we show a defocusing situation:
it is worthwhile to note that the formula defining the angle 
$\alp_{2}$ is given by the same $x-$derivative
of  geodesic calculated in the source coordinates. Thus, the relative
change of luminosity in the planar configuration is equal to
\beq
\label{3.7}
\frac{l_{2}-l_{1}}{l_{1}}=\frac{\alp_{2}-\alp_{1}}{\alp_{1}}\,.
\eeq
We have either an increasing luminosity  for
$\alp_{2}>\alp_{1}$,
(focusing case) or a decreasing one for
$\alp_{2}<\alp_{1}$ (defocusing case).
Actually, to obtain 
an observable change of luminosity, we have to consider the square of 
(\ref{3.7}) since we have
to take into account solid angles in the space, that is
\beq
\label{3.10}
\frac{L_{2}-L_{1}}{L_{1}}=\frac{\Delta L}{L}=
\pm\left(\frac{\alp_{2}-\alp_{1}}{\alp_{1}}\right)^{2}\,.
\eeq
Plus sign corresponds to the focusing situation, minus sign 
to the defocusing one. A general formula for relative
change of luminosities is
\beq
\label{3.11}
\frac{\Delta L}{L}=
\pm\left[\left(\frac{dy_{2}(x)}{dx}-
\frac{dy_{1}(x)}{dx}\right)
\left(\frac{dy_{1}(x)}{dx}\right)^{-1}\right]_{({x_{S}},{y_{S}})}^{2}\,,
\eeq
where
the two  derivatives of geodesics are calculated in the coordinates of the 
source.

Let us now apply these general considerations
to the case of a flat metric which is
perturbed by the gravitational field of a moving MACHO.
This means that the initial metric $g^{(1)}\dmunu$ is a 
 Minkowski one while the  metric $g^{(2)}\dmunu$ is
 a Schwarzschild  one.
Without MACHO, geodesics are straight lines emitted by the source (Fig.1),
that is 
\beq
\label{3.14}
r=\frac{r_{0}}{\cos(\p-\p_{0})}\,,
\eeq
in polar coordinates, or
\beq
\label{3.15}
r_{0}=Ax+By\,,
\eeq
in Cartesian coordinates. The constants $A$ and $B$ are the same 
as (\ref{2.20}).

When a MACHO perturbs the flat background, the geodesics 
 are given by Eq.(\ref{2.10}) (or (\ref{2.17})).

Let us now consider the simplest case of a MACHO of 
mass $M$ posed in the origin of coordinate.
In the focusing situation (see Fig.2),
 the lens
(a MACHO with Cartesian coordinates
$\{0,0\}$) is between the stellar source (posed in $\{-a,0\}$) 
and the observer
(with the aperture of telescope in $\{R,h\}$). 
We calculate, in the source coordinates, 
the derivative of light ray trajectory (\ref{3.14}) 
or (\ref{3.15}), when the MACHO
is not present obtaining
\beq
\label{3.16}
\frac{dy_{1}}{dx}\left|\right._{(-a,0)}=-\left(\frac{A}{B}\right)\,,
\eeq
and when it is present (by using (\ref{2.10}) or (\ref{2.17}))
\beq
\label{3.17}
\frac{dy_{2}}{dx}\left|\right._{(-a,0)}=-\left\{\frac{A+
\left(\frac{R_s}{2r_{0}}\right)(A^{2}-2)}{B\left[1+
A\left(\frac{R_s}{r_{0}}\right)\right]}\right\}\,.
\eeq
From (\ref{3.10}) (or (\ref{3.11})), 
we immediately obtain the relative change of
luminosity induced on the telescope, that is
\beq
\label{3.18}
\frac{\Delta L}{L}
=\left(\frac{R_s}{2r_{0}}\right)^{2}
\left\{\frac{A^{2}+2}{A\left[1+
A\left(\frac{R_s}{r_{0}}\right)\right]}\right\}^{2}\,.
\eeq
For defocusing, as shown in Fig.3,
the source is in $\{a,0\}$.
Performing the same calculation as above, we get
\beq
\label{3.16'}
\frac{dy_{1}}{dx}\left|\right._{(a,0)}=-\left(\frac{A}{B}\right)\,,
\eeq
when the MACHO is not present and
\beq
\label{3.17'}
\frac{dy_{2}}{dx}\left|\right._{(a,0)}=-\left\{\frac{A-
\left(\frac{R_s}{2r_{0}}\right)(A^{2}-2)}{B\left[1-
A\left(\frac{R_s}{r_{0}}\right)\right]}\right\}\,,
\eeq
when it is present.
The relative change of
luminosity induced on the telescope is
\beq
\label{3.18'}
\frac{\Delta L}{L}
=-\left(\frac{R_s}{2r_{0}}\right)^{2}
\left\{\frac{A^{2}+2}{A\left[1-
A\left(\frac{R_s}{r_{0}}\right)\right]}\right\}^{2}\,.
\eeq
It is worthwhile to note that if 
${\displaystyle A\simeq \frac{r_{0}}{R_{s}}}$
in Eq.(\ref{3.18'}), the relative change can be huge.
The couples of Eqs. (\ref{3.17}), (\ref{3.18}) and 
(\ref{3.17'}), (\ref{3.18'}) show that the variation of luminosity strictly
depends on  the relative 
positions of the MACHO (lens)  and 
the ray ($r_{0}$ and $\p_{0}$),  on the mass of the MACHO (to be
more precise on the product $GM$ where the gravitational coupling $G$ is 
supposed constant) and on the relative position of the lens and the
light source (the signs inside Eqs.(\ref{3.18}) and (\ref{3.18'})
 depend on taking $x_{S}=-a$ or $x_{s}=a$ and show that the problem
of focusing and defocusing is not completely symmetric). 
 Such calculations can be performed in any  position of source
and  lens, here, for simplicity, we have taken into account source, lens
and observer which lie on the same line. These results can be used
to estimate the quantities of microlensing theory as we shall do
in next section.

\section{\normalsize \bf The mass of MACHO and the optical depth }

First of all, by the above formulas, we can  calculate the mass 
of  MACHO (lens)
both for focusing and defocusing cases; in fact from Eq.(\ref{3.18})
and (\ref{3.18'}),
using (\ref{schwarzschild}), we obtain
\beq
\label{3.21}
M=\left(\frac{c^{2}r_{0}}{G}\right)
\left[\frac{A\sqrt{|\Delta L/L|}}{2+
A^{2}\mp 2A^{2}\sqrt{|\Delta L/L|}}\right]\;.
\eeq
Now minus sign refers to focusing and plus to defocusing; furthermore, we 
are considering the absolute relative variation of light intensity,
in the sense that, given a luminosity curve of a source, both the peak
or the valley can give indications on the MACHO mass.

We have to stress that $ M=M(\Delta L/L, r_{0}, \p_{0})$,
that is, in principle, we can get the mass of  the MACHO  only
by knowing its position with respect to the source $\{r_{0},\p_{0}\}$ 
and the relative
variation of intensity. The term in $A^{2}=\cos^{2}\p_{0}$
and $\sqrt{|\Delta L/L|}$ at the denominator tells us whether the 
deflecting body is between the light source and the observer or
behind the light source.

By using Eqs.(\ref{0.14}) and (\ref{optical}), the optical depth is
given by
\beq
\label{depth}
\tau_{\pm}= \left(\frac{r_{0}D_{ls}}{D_{ol}^{2}}\right)
\left[\frac{A\sqrt{|\Delta L/L|}}{2+
A^{2}\mp 2A^{2}\sqrt{|\Delta L/L|}}\right]\;,
\eeq
where $\tau_{+}$ is the optical depth (probability) connected to a focusing
event while $\tau_{-}$ is associated to a defocusing one.
If $r_{E}=\theta_{E}D_{ol}$, the direct dependence on $\theta_{E}$
appears in (\ref{depth}). It is worthwhile to note that now the 
distances $D_{ls}, D_{os}, D_{ol}$ explicitely give a contribution
telling us that optical depth (\ie the probability to obtain lensing events)
strictly depends on the geometry of the optical system 
source--lens--observer.

Another important issue is the duration of the relative
luminosity variation of the source.
If we consider $r_{0}\simeq r_{E}$ and approximate
Eqs.(\ref{3.18}) and (\ref{3.18'})
in terms of ${\displaystyle \left(\frac{R_s}{r_{0}}\right)}$, 
we get
\beq
\label{4.1}
\frac{\Delta L}{L}\simeq\pm\left(\frac{\tilde{A}R_s}{v\Delta t}\right)^{2}\,,
\eeq
where ${\displaystyle \tilde{A}=\left(\frac{A^{2}+2}{A}\right)^{2}}$.
From (\ref{4.1}), it is easy to see that the luminosity 
variation strictly depends
on the velocity of a passing MACHO and on the time in which it remains
in the Einstein ring. A fast passing MACHO will produce a sharp
peak (or valley)  of luminosity.

In order to give some estimation let us
consider Eq.(\ref{3.21}): we  obtain a MACHO of mass 
$M\sim 0.5\div 1 M_{\odot}$, 
if ${\displaystyle \frac{\Delta L}{L}\sim 10^{-2}}$,
$r_{0}\simeq r_{E}\sim 1$AU and 
${\displaystyle \p_{0}\sim \frac{\pi}{2}+|\delta|}$,
with ${\displaystyle |\delta|\sim 10^{-5}}$. Such result holds for
focusing and defocusing MACHOs. On the other hand, it is easy to 
obtain optical
depth of the order $\tau\sim 10^{-6}$ as determined by
the OGLE and MACHO collaborations toward the Galactic bulge
\bib{paczynski},\bib{alcock}, or $\tau\sim 10^{-7}\div 10^{-8}$ 
as estimated toward the LMC \bib{alcock1}. 
It is interesting to see that in this second case the ratio 
${\displaystyle \frac{D_{ls}}{D_{os}}}$ is close to the 
unity since the distances in the 
Galactic halo and in LMC are similar, so being $r_{0}\sim r_{E}$,
$\tau$ depends only on the two angles $\theta_{E}$, $\p_{0}$ and on the
relative luminosity variation.
The same results
are also obtained for 
if ${\displaystyle \frac{\Delta L}{L}\sim 10^{-4}}$ and
${\displaystyle |\delta|\sim 10^{-3}}$. 

In principle, we can cover all
the potentially detectable mass range $10^{-6} M_{\odot}$ 
to $10^{2} M_{\odot}$    expected for MACHOs.

 However,  we can use also the
relative source--lens velocity $v$  
and the duration of luminosity variation $\Delta t$. The above mass 
$M\sim M_{\odot}$ is detected for $v\sim 200$ Km s$^{-1}$ and 
$\Delta t\sim 10^{6}$s.

This second method is good for measurements inside the Galaxy since
the velocities are quite well known \bib{binney} and the distances 
$r_{0}\sim r_{E}=r_{E}(D_{l},D_{ls},D_{s})$ can be accurately estimated.

\section{\normalsize \bf  Conclusions}
In this paper we have pointed out that  we can get a microlensing effect
not only if we detect an increasing luminosity for a given source, but
also if we detect a decreasing one. 
Furthermore, by the  knowledge of  the geometry of the  
system source--lens--observer,
we can estimate both mass and optical depth of a given lens.
 These facts could contribute to bypass one of the lack of
microlensing detecting experiments: the low number of observed events
(till now about 100,  not all exactly tested, 
for millions of detected source stars).
Roughly speaking,
 one could
expect to double the number of succesful detections including
also defocusing events.

It is worthwhile to note that when several MACHOs are present, the 
previous discussion still holds due to the Fermat principle (see, for example
\bib{ehlers}). In fact, any compact object perturbs the flat 
gravitational background and a light ray passing through the locally
perturbed metrics $g^{(2)}\dmunu, \cdots, g^{(j)}\dmunu$ undergoes 
$j-1$ deviations. The effect is additive and it is similar to that of a 
light ray passing
through different media with refraction indexes $n_{1},\cdots n_{j}$.
Then, in principle, it is possible to evaluate the total deviation of
a light ray  by summing up the effects 
of the various deflectors.

We have to stress that in a statistical approach to the
microlensing, 
we have two contributions to the
number density $n(D_{l})$ of lenses, one coming from focusing objects
$n_{+}(D_{l})$ and another coming from defocusing ojects $n_{-}(D_{l})$. 
As a final remark, we would like to stress that the approach we have 
developed in this paper could be useful to reconsider some faint sources
which are expected to be brighter because of their mass.

In a forthcoming paper, we will develop such a statistical approach 
considering focusing and defocusing lenses and
giving workable models which can be used in the measurements
toward the bulge of  Galaxy or toward the LMC.

\vspace{4. mm}

{\bf ACKNOWLEDGMENTS}\\
The authors want to thank Michal Jaroszy\'{n}ski for the useful
discussions on the topic. V.I.M. thanks Prof. M. Capaccioli
for the hospitality  in Osservatorio Astronomico di Capodimonte.

\begin{center}
{\bf REFERENCES}
\end{center}

\begin{enumerate}
\item\label{schneider}
P. Schneider {\it Cosmological Applications of Gravitational Lensing},
Lecture Notes in Physics, eds. E. Martinez--Gonzales, J.L. Sanz,  
Springer Verlag, Berlin (1996), Astro--Ph/9512047 (1995).
\item\label{ehlers}
P.V. Blioh and A.A. Minakov {\it Gravitational Lenses}
Kiev, Naukova Dumka (1989) (in Russian).\\
P.Schneider, J. Ehlers, and E.E. Falco {\it Gravitational Lenses}
Springer--Verlag, Berlin (1992).\\
R. Kaiser {\it Gravitational Lenses} 
Lecture Notes in Physics {\bf 404}, p. 143, Springer--Verlag, Berlin (1992). 
\item\label{peebles} 
P.J.E. Peebles  {\it Principles of 
Physical Cosmology} (Princeton Univ. Press.,
Princeton 1993)
\item\label{paczynski} 
B. Paczy\'{n}ski, \apj {\bf 301}, 503 (1986);\\
B. Paczy\'{n}ski, \apj {\bf 304}, 1 (1986);\\
B. Paczy\'{n}ski {\it Gravitational Lenses} 
Lecture Notes in Physics {\bf 406}, p. 163, Springer--Verlag, Berlin (1992).
\item\label{padmanabhan}
T. Padmanabhan, {\it Structure Formation in the Universe},
Cambridge Univ. Press, Cambridge (1993).
\item\label{gurevich1}
A.V. Gurevich, V.A. Sirota, and K.P. Zybin, \pl {\bf 207 A}, 333 (1995).
\item\label{gurevich2}
A.V. Gurevich K.P. Zybin, \pl {\bf 208 A}, 276 (1995).
\item\label{gurevich3}
A.V. Gurevich,  K.P. Zybin,  and V.A. Sirota, \pl {\bf 214 A}, 232 (1996).
\item\label{carr}
B.J. Carr, in {\it Proceedings of Int. Workshop on the Identification
of Dark Matter}, Sheffield 8--12 Sept. 1996 (World Scientific).
\item\label{chang}
K. Chang and S. Refsdal, {\it Nature} {\bf 282}, 561 (1979).
\item\label{irwin}
M.J. Irwin et al. \aj {\bf 98}, 1989 (1989).
\item\label{schild}
R.E. Schild and R.C. Smith, \aj {\bf 101}, 813 (1991).  
\item\label{alcock}
C. Alcock et al., {\it Nature} {\bf 365}, 621 (1993). 
\item\label{shapiro}
I.I. Shapiro, \prl {\bf 13}, 789 (1964).
\item\label{aubourg}
E. Aubourg et al., {\it Nature} {\bf 365}, 623 (1993).
\item\label{udalski}
A. Udalski et al., {\it Acta Astron.} {\bf 42}, 253 (1992).
\item\label{alard}
C. Alard in: {\it Astrophysical Applications of Gravitational Lensing}
Proc. IAU Symp. (1995)
173, eds. C.S. Kochanek and J.N. Hewitt (Boston, Kluwer)
\item\label{papapetrou}
A. Papapetrou, {\it Lectures on General Relativity}, Reidel Pub. Company,
Boston (1974).
\item\label{ellis}
S.W. Hawking and G.F.R. Ellis, {\it The Large Scale Structure of 
Space--Time}, Cambridge Univ. Press (1973) Cambridge.
\item\label{harwit}
M. Harwit, {\it Astrophysical Concepts}, Ed. Springer, Berlin (1988).
\item\label{refsdal}
S. Refsdal, {\it Proc. Intern. Conf. on Rel. Theories of Gravitation},
London (1970).\\
S.Refsdal, \apj {\bf 159}, 357 (1970).\\
S. Refsdal and J. Surdej, {\it Rep. Prog. Phys.} {\bf 57}, 117 (1994).
\item\label{press}
W.H. Press and J.E. Gunn, \apj {\bf 185},  397 (1973).
\item\label{wald}
R.M. Wald, {\it General Relativity}, Univ. of Chicago Press, Chicago (1984).
\item\label{byrd}
P.F. Byrd and M.D. Friedman {\it Handbook of elliptical integrals...},
Springer--Verlag, Berlin (1954).
\item\label{alcock1}
C. Alcock, R.A. Allsman, T.S. Axelrod et al., \apj {\bf 461}, 84 (1996).
\item\label{binney}
J. Binney and S. Tremaine, {\it Galactic Dynamics}, Princeton Univ. Press,
Princeton (1987).
\end{enumerate}
\vfill

\newpage
\begin{center}
{\bf FIGURE CAPTIONS}
\end{center}
\noindent {\bf Fig.1} Schematic representation of the system 
source--observer in a Cartesian reference frame. In this case the lens is not
present and the metric $g^{(1)}\dmunu$ can be assumed to be  
Minkowski. The source is in $\{-a, 0\}$, the upper edge of the telescope
(collecting the last light ray) is in $\{R, h\}$.
\vspace{2. mm}

\noindent {\bf Fig.2} As above, but now a MACHO (lens) is present in the 
origin of coordinates. The metric is  $g^{(2)}\dmunu$ which is the
Schwarzschild local perturbation of $g^{(1)}\dmunu$. This is a focusing
situation since the telescope collects more light than above
(\ie more geodesics, due to the action of the lens in the origin,
converge in the telescope).  
\vspace{2. mm}

\noindent {\bf Fig.3} The defocusing situation. The positions of the source
and of the lens are inverted with respect to the observer. The telescope
collects less light than in the case shown in Fig.1 (and obviously in the
case in Fig.2).
\vspace{2. mm}

 \vfill

\end{document}